\documentclass[letterpaper,twocolumn,conference]{IEEEtran}
\usepackage[T1]{fontenc}
\usepackage[latin9]{inputenc}
\usepackage{amstext}
\usepackage{graphicx}

\makeatletter

\pdfpageheight\paperheight
\pdfpagewidth\paperwidth

\usepackage[caption=false,font=footnotesize]{subfig}
\usepackage{cite}
\usepackage[hang,flushmargin]{footmisc}

\newcommand\blfootnote[1]{%
  \begingroup
  \renewcommand\thefootnote{}\footnote{#1}%
  \addtocounter{footnote}{-1}%
  \endgroup
}

\@ifundefined{showcaptionsetup}{}{%
 \PassOptionsToPackage{caption=false}{subfig}}
\usepackage{subfig}
\makeatother

\begin{document}

\title{The Socket Store: An App Model for the Application-Network Interaction}

\author{Christos Liaskos, Ageliki Tsioliaridou, Sotiris Ioannidis\\
{\small{}Foundation for Research and Technology - Hellas (FORTH)}\\
{\small{}Emails: \{cliaskos,atsiolia,sotiris\}@ics.forth.gr}}
\maketitle
\begin{abstract}
A developer of mobile or desktop applications is responsible for implementing
the network logic of his software. Nonetheless: i) Developers are
not network specialists, while pressure for emphasis on the visible
application parts places the network logic out of the coding focus.
Moreover, computer networks undergo evolution at paces that developers
may not follow. ii) From the network resource provider point of view,
marketing novel services and involving a broad audience is also challenge
for the same reason. Moreover, the objectives of end-user networking
logic are neither clear nor uniform. This constitutes the central
optimization of network resources an additional challenge. As a solution
to these problems, we propose the Socket Store. The Store is a marketplace
containing end-user network logic in modular form. The Store modules
act as intelligent mediators between the end-user and the network
resources. Each module has a clear, specialized objective, such as
connecting two clients over the Internet while avoiding transit networks
suspicious for eavesdropping. The Store is populated and peer-reviewed
by network specialists, whose motive is the visibility, practical
applicability and monetization potential of their work. A developer
first purchases access to a given socket module. Subsequently, he
incorporates it to his applications under development, obtaining state-of-the-art
performance with trivial coding burden. A full Store prototype is
implemented and a critical data streaming module is evaluated as a
driving case. \blfootnote{This work was funded by the European Union via the Horizon 2020: Future Emerging Topics call (FETOPEN), grant EU736876, project VISORSURF (http://www.visorsurf.eu).}
\end{abstract}


\IEEEpeerreviewmaketitle{}

\section{Introduction\label{sec:Introduction}}

\IEEEPARstart{I}{ndustrial} and academic research in computer networks
has bloomed in the past decades. Sub-disciplines have been defined,
and state-of-the-art solutions have been proposed for almost every
conceivable topic. A huge knowledge potential has been accumulated,
which is kept up to date with scientific breakthroughs, culminating
with the recent advancements in virtualization and Software-Defined
Networking (SDN). The interested networks expert can use mature classification
systems provided by formal publications bodies, become current with
latest advancements and innovate in platforms that are more than ever
modular and hardware-decoupled.

Switch you point of view to that of the average developer of mobile
or desktop applications. You have a promising idea for a new \emph{app}
that will provide a novel utility in every-day life. You need a graphical
interface that stands out and multi-platform support, whilst being
pressed for minimal development time. What part of your coding effort
will be devoted to ``under-the-hood'' aspects, i.e., to proper network
programming? Moreover, how accessible is the network research knowledge
pool to you? Assuming that the proper time is invested, will you also
consider the newest network infrastructure capabilities or reported
issues?

Expectedly, the problem of unused research knowledge potential exists
and has important consequences. Specifically, surveys identify the
inefficient app-side communication programming as the top factor impacting
both application and network performance~\cite{Balachandran.2013,acmSoftwareQuality.2014}.
Moreover, in a 2016 study, a $66\%$ percentage of queried professionals
acknowledged difficulty in keeping up with the advances in the networking
field~\cite{Loeb.2016}. These studies exhibit: i) a clear lack of
networking code sophistication during app development, and ii) unwillingness
to overcome this problem via educative means or additional development
effort. A partial solution to the problem are client-side libraries
that have appeared, which expose simple programming interfaces and
hide the networking complexity~\cite{Evensen.2015}. Despite their
high importance, such solutions seek to provide basic connectivity,
and not research state-of-the-art performance or advanced functionality.
Moreover, given that they constitute app-side solutions, the goal
of close interaction between applications and the network remains
elusive at large, despite the rapid advances in related infrastructure
capabilities.

From the network resource provider aspect, inefficient client-side
networking code can also be problematic. Without close interaction
with the app, the network must strive to equally serve both sane and
insane app-side logic. The latter may comprise buggy, greedy or malevolent
behavior, allowing for, e.g., DoS attacks~\cite{INFOCOM,CCR}. Moreover,
sophisticated services offered by network and resource providers,
such as Infrastructure-as-a-Service, virtualized functions and protocol
stacks, require a good degree of specialization~\cite{ETSI.2013,Akamai.Tough.2016},
limiting their use to well-trained network specialists. Companies
are already seeking the involvement of people from the app development
world, in order to increase the commercial exposure of such services
and encourage their further evolution~\cite{Akamai.Markeplace.2016,CanonicalLtd.2016}.

The present work contributes the Socket Store, a novel way for connecting
the research knowledge potential to the software development world.
The Store is platform and marketplace involving network specialists,
developers and network resource providers, each with well-defined,
fitting roles and motivation. Within the Store, network resource providers
expose the infrastructure capabilities, and specialists publish reusable,
network logic modules that operate on top of them. Developers browse
the Store, purchase access to modules fitting their application, and
simply invoke them with minimal coding~\cite{Stevens.2004}. Internally,
the Store uses a universal and natural modeling approach: Multi-agents
are employed to model all network resources, network logic modules,
the applications and all their cross-interactions. Through the Store
specialists gain visibility, automatic and fair performance evaluation
of their contributions, as well as monetization potential of their
research. Network resource providers gain a clear, uniform understanding
of the preferences of their end-clients, network optimization potential
and better commercial visibility of advanced services. Finally, the
developers can minimize their expenses by cutting down development
time, while obtaining network performance at the level of research
state-of-the-art.

The paper is organized as follows. Section~\ref{sec:System-Overview}
provides a high-level view of the Store workflow. Section~\ref{sec:Multi-Agent-Systems}
gives an introduction to multi-agent systems. The Store architecture
is detailed in Section~\ref{sec:The-Socket-Store-Arch}. A full prototype
of the Store is implemented and evaluated in an SDN environment in
Section~\ref{sec:Implementation}. Section~\ref{sec:Summary-and-Outlook}
further discusses what the Store can offer and what it does not. The
conclusion is given in Section~\ref{sec:Conclusion}.

\section{System Overview\label{sec:System-Overview}}

The Socket Store is a market place where software developers without
specialty on computer networks can purchase intelligent network code.
This code is intended to offer high performance and serve specialized
tasks, while necessarily being easy to use and well-tested. Thus,
the Store provides:
\begin{itemize}
\item A uniform and natural model to represent the interaction between an
application and the network. An agent-based modeling is employed,
which is generic and hides the network-related specifics from the
developer.
\item A robust code review process, which is strongly tied to the academic
research evaluation process.
\item Access to testbeds for automatic and fair network code evaluation
and ranking.
\end{itemize}
The Store refers to the interaction of software applications with
existing network resources, prioritizing compatibility. However, it
allows network resource provides to publish novel, specialized services
as well, promoting the evolution of networking.

\subsection*{Involved Parties}

The Store involves three categories of professionals. The Software
Developers, the Network Specialists and the Network Resource Providers.

The first group (Developers) refers to any individual or company that
develops software in the broad sense. Emphasis is placed on professionals
that develop end-user mobile or desktop applications. In many such
cases, where an end-user interacts directly with the software, it
can be more significant for the developer to focus his efforts on
the end-user's visual and auditory experience. Under-the-hood concerns,
such as optimizing the interaction between the application and the
network resources, can constitute a considerable overhead in terms
of development time and investment in obtaining the required know-how.

The Network Specialists group represents the individuals or companies
that specialize in developing networking solutions. The group can
comprise academic and industrial research teams, which see interest
in pinpointing network performance bottlenecks or opportunities, performing
thorough studies of related work, and providing state-of-the-art contributions.
All areas of specialty that can be practically applied are considered,
exemplary including quality of service, security, energy efficiency
and resource sharing.

The Network Resource Providers is a broad term comprising privately
help companies or the public, offering physical, virtual or logical
resources. Companies may specialize in Cloud provision, secure overlay
networking, Internet state monitoring services via passive/active
measurements and more. Providers can choose to actively participate
to the Socket Store, facilitating the use of resources by the Specialists~\cite{Akamai.Markeplace.2016}.
However, this approach is not mandatory since the Specialists may
themselves model and use existing resources and services.

Finally, the Socket Store is transparent to plain users, who download
and use applications normally.

\subsection*{Workflow}

\begin{figure}[t]
\begin{centering}
\textsf{\includegraphics[width=0.9\columnwidth]{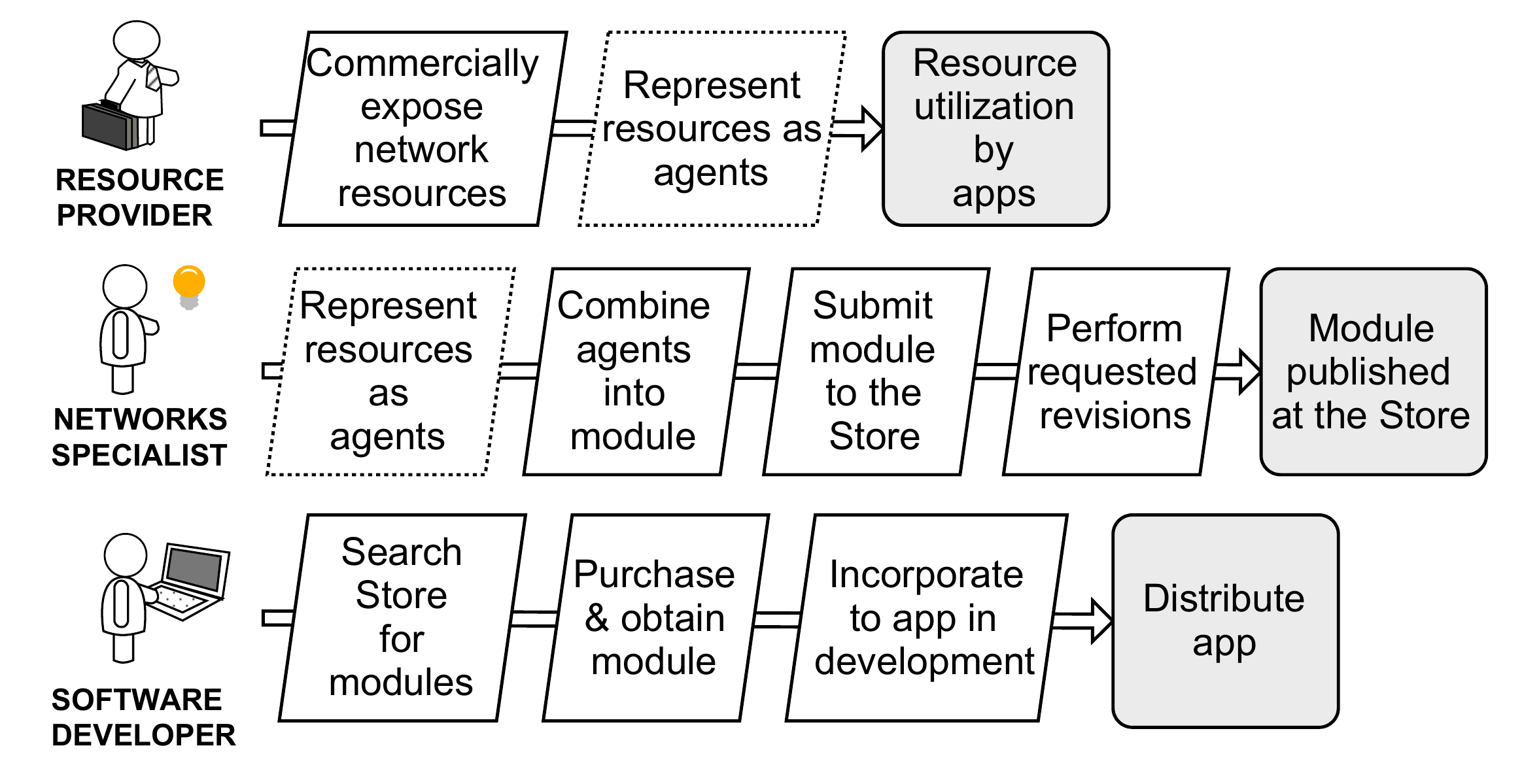}}
\par\end{centering}
\caption{\label{fig:OVERVIEW}Overview of the Socket Store workflow for each
involved party. }
\end{figure}
The workflow involving the Store and each party is shown in Fig.~\ref{fig:OVERVIEW}.
The Developers are considered to be in the process of creating or
maintaining a network-enabled application. Instead of programming
the interaction with the network themselves, Developers browse the
Store for network modules that fit their needs. Subsequently, they
purchase and incorporate it to the developed application, and proceed
to push it to their distribution channels. The term ``Purchase''
has the same meaning as in app stores: obtaining for a monetary price
or for free. Enabling modules as in-app purchases is also allowed.

The Specialists are responsible for creating innovative and competitive
Socket modules. To this end, they combine existing resource agents
or create new agent-based representations of resources. A module undergoes
a submission, followed by a cycle of reviewing by other researchers,
and revisions performed by the original module creators. Moreover,
each submitted module is required to be associated with one or more
Store performance metrics, in order to be automatically evaluated
in testbeds and ensure its conformity with the promised functionality.
Upon successful revision, a module is accepted for publication at
the Store, becoming browse-able and usable by Developers.

The workflow of Resource providers is not necessarily altered. The
providers continue to normally offer access to their resources, without
being obliged to offer corresponding agent-based representations;
the Specialists can perform this task independently. Nonetheless,
extra visibility via the Store can increase the utilization of resources,
motivating the providers to provide the corresponding agent representations
directly.

Given the central placement of agent-based modeling in the Store,
we proceed to provide the necessary background in the next Section.

\section{Multi-Agent Systems\label{sec:Multi-Agent-Systems}}

\begin{figure}[tp]
\begin{centering}
\textsf{\includegraphics[width=0.9\columnwidth]{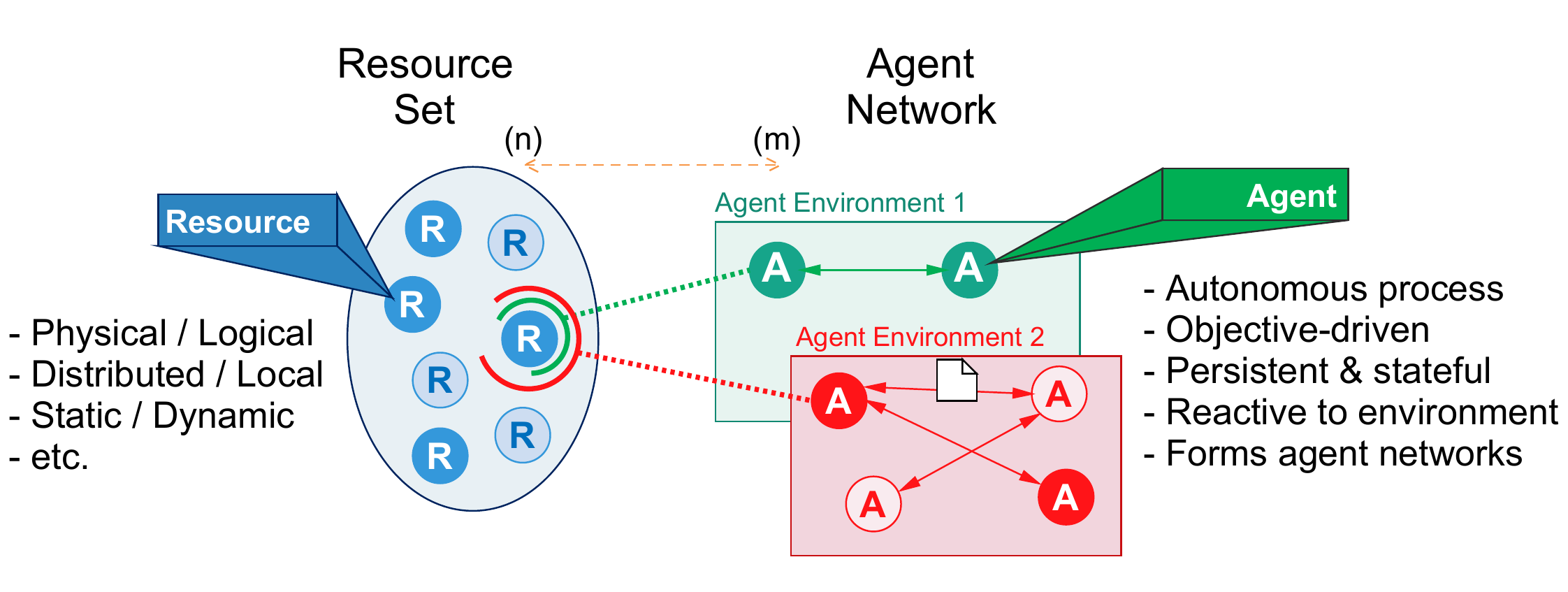}}
\par\end{centering}
\caption{\label{fig:agents}Schematic of an agent-based representation of a
set of arbitrary resources. An agent environment represents a concern
context. An agent in such an environment represents one or more resources
in the given context. }
\end{figure}
The agent-based modeling is an approach for representing and managing
an arbitrary set of resources in a natural and intuitive manner. The
term resource is generic and can cover any kind of physical or logical,
distributed or local, static or dynamic potential~\cite{chang2015multi}.
Diverse examples include physical processors, link capacity monitors
and data flow routing services within a computer network.

The management of resources commonly presents cross-cutting concerns:
one management aspect may exemplary refer to the energy consumption
of a physical computer network, while another may refer to the efficiency
of the logical path allocation services that run within it. In agent-based
modeling, such concerns are denoted as Environments, as shown in Fig.~\ref{fig:agents}.
Each environment acts as a unified context for autonomous software
processes, the Agents.

An agent represents and manages one or more resources within an environment.
For instance, an agent may represent and manage the routing table
of a single router, or the peering relations of a complete autonomous
system, depending on the required level of abstraction. An agent manages
its associated resource autonomously, i.e., without the need for constant
human supervision, towards a clearly defined Objective. Returning
to the routing table example, an agent's objective can be to find
and maintain a list of the shortest network paths within a hybrid
network comprising SDN and legacy parts. In order to accomplish their
objective, agents commonly collaborate with each other and react to
global changes in their environment. Additionally, they can move within
their environment to optimize their objective: An agent representing
a virtual machine can relocate to another physical site that offers
better performance.

The agent-based approach presents enticing features that constitute
it an ideal platform for implementing the Socket Store. From a practical
point of view, agents and environments are programmatically described
by customizable object classes, which are the norm in software development.
Moreover, existing agent development frameworks provide: i) Agent
management facilities comprising objective definition, instantiation,
inter-agent communication, movement and destruction. ii) Environment
synchronization, allowing agents to obtain a central and coherent
view of their environment, even when the represented resources are
purely distributed. This generality constitutes multi-agent systems
capable of efficiently representing both distributed \emph{and} centralized
network architectures, such as SDNs. iii) GUIs for the visualization
of the inspection of agents and their environments. iv) Cross-platform
operation regarding programming languages for describing the agents
and the hardware/software platforms they can reside in. Additional
benefits include massive scalability and strong community support~\cite{chang2015multi}.

\section{The Socket Store Architecture\label{sec:The-Socket-Store-Arch}}

\begin{figure*}[tbph]
\begin{centering}
\textsf{\includegraphics[bb=0bp 15bp 852bp 330bp,clip,width=0.9\textwidth,height=5.5cm]{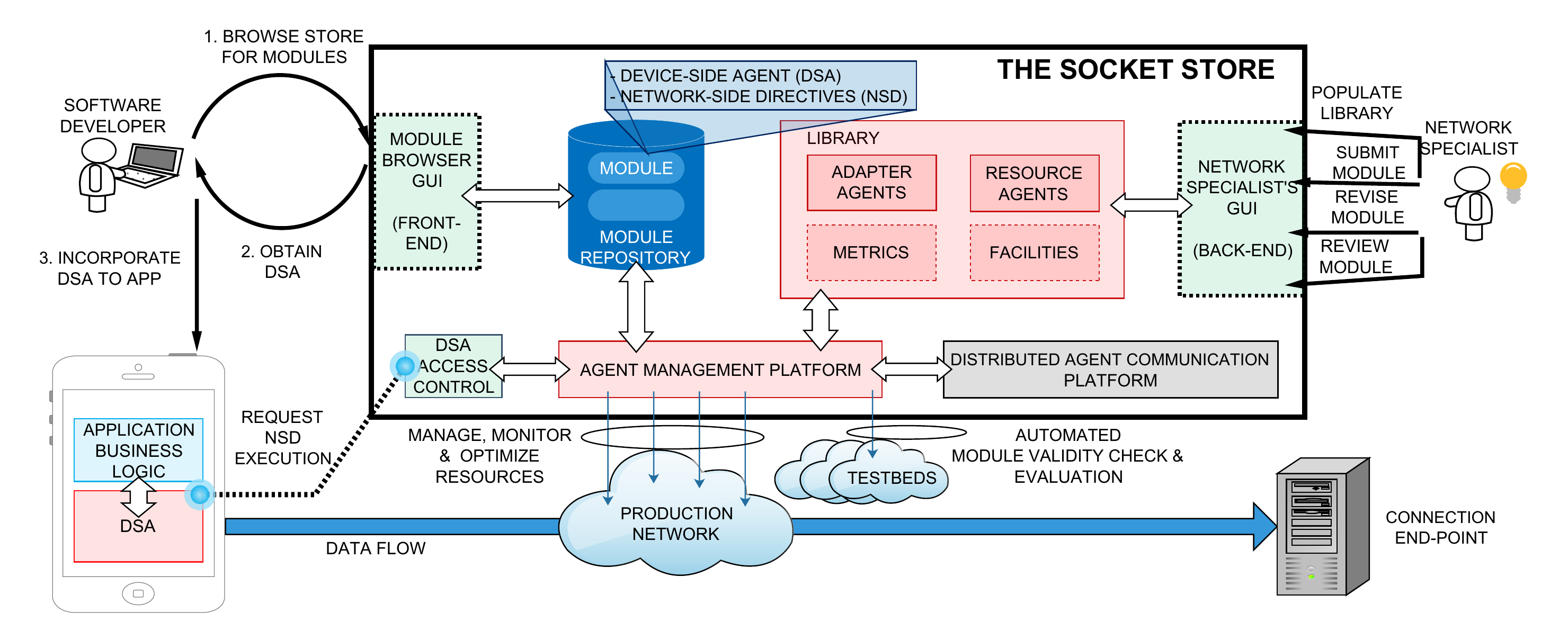}}
\par\end{centering}
\caption{\label{fig:DETAILED}Architecture of the Socket Store, showing its
main components, the Developer and Specialist's interactions, and
the end-device incorporation approach.}
\end{figure*}
The architecture of the Socket Store comprises (Fig.~\ref{fig:DETAILED}):
(i) A front-end and a back-end interface, for Developers and Specialists
respectively, (ii) The Agent Type Library, (iii) The Module Repository,
(iv) The Store Access Control, (v) The Agent Instance Management Platform,
operating in tandem with a Distributed Agent Communication platform.
Finally, closely related but external to the Store are the Testbeds,
which allow for automated module validity check and evaluation when
present.

The Specialists employ the Store's back-end to submit, review, revise
and maintain modules. A module is an XML description of a set of agents
which are created, placed, moved, monitored and destroyed per the
instructions, i.e., the Network Side Directives (NSD) of the Specialist
authoring the module. A module is associated with a performance Metric,
which expresses its promised objective. The agent types and the metrics
are organized as the Agent Type Library of the Store. Within the Library,
the agent types are further classified as Resource Agents, which directly
model a network resource, or as Adapters, which combine other agents
to provide a different modeling approach or functionality.

A module also contains a Device-Side Agent (DSA), an \emph{open-source},
lightweight agent intended to be executed at the end-user device,
and perform access control to the Store and, subsequently, to network
resources. The Store provides standard DSAs for different end-device
operating systems. An important but optional capability, is that specialists
can extend these standard DSAs to offer advanced interaction between
the application and the network resources. An example is given in
Section\ \ref{sec:Implementation}. Modules accepted for publication
at the Store have their descriptions and DSAs stored centrally at
the Module Repository.

The Developers use the Store's front-end interface to search for and
purchase modules fitting the requirements of their applications. The
front-end is intended to offer a smart search engine, a module ranking
and a commenting system. Moreover, each module can be accompanied
by its associated Metric values, logged during its operation. Thus,
an objective module ranking can be attained apart from a community-based
one. Once a module has been obtained, the Developer incorporates the
DSA to his application, with trivial coding burden.

An end-user that executes the application on his device will trigger
the execution of the DSA. The DSA then undergoes authorization by
the Store Access Control, described later in Section~\ref{subsec:Security-Model},
ensuring that the application has indeed access to execute the associated
NSD of the module. Upon successful authorization, the Agent Instance
Management Platform executes the NSD creating the respective agents
within their environments, which in turn allocate the corresponding
network resources. While the module is still in review phase, the
NSD is executed within testbeds, providing automated module validity
check and performance evaluation. Published modules are ready for
deployment in their intended production networks, which can include
the Internet.

The Distributed Agent Communication platform supports the Agent Instance
Management, providing robust inter-agent and agent-Store communication.
Existing distributed messaging solutions, such as ZeroMQ~\cite{Akgul:2013:ZER:2523409},
fill in this role. The Agent Instance Management is thus enabled to
provide a central view of the active agents and their resources, regardless
of the actual distribution degree of the network. Thus, the Store
attains a unified modeling and management of both completely centralized
network technologies, such as SDNs, or completely decentralized. The
centralization benefits the Specialists and the Resource Providers:
it can facilitate the development and debugging of modules, as well
as the optimization of network resource allocation.

\subsection{Security and Trust Model\label{subsec:Security-Model}}

The Socket Store aligns itself with existing, deployed and well-received
security and trust approaches, without introducing Store-specific
novelties. Particularly, it leverages existing application distribution
channels, resource slicing and control/data Separation as follows:

Socket Store modules are distributed as parts of standard applications
and not in a stand-alone fashion. Thus, they employ the same privacy
agreements and trust levels of the corresponding application distribution
channels. Distribution via App Stores also solves the problem of access
control. Upon DSA invocation, the Socket Store contacts the App Store
to verify ownership of the \emph{application} containing the module.
The same, existing approach is employed when modules are provided
as in-app purchases. Distribution outside App Stores relies on any
standard licensing approach, such as existing license server schemes.
Notice that the DSAs are supplied as open-source in any case, to ensure
the Developer's trust.

Once the NSD part of a module is executed, the instantiated agents
call upon existing ways of interacting with resources offered by the
Providers~\cite{Akamai.Tough.2016}. This approach does not upset
the existing trust model associated with the use of such resources.
Additionally, existing sand-boxing techniques, such as network resource
slicing remain active and unobstructed by the Store, ensuring that
programming faults or malevolent behavior does not affect the network
as a whole.

A key-point in the Socket Store trust model is that it operates at
the network control plane. Via their modules, Specialists propose
optimizations in the management of resources, without access to the
actual data flowing through the network. As a step against indirect
security breaches, all agent definitions are required to rely on the
Store Agent Library only, and not on custom libraries.

The Store logs all agent actions to trace errors or malevolent behavior.
Moreover, all module submissions are made by eponymous Specialists,
who are encouraged to extend the prestige of their professional profile
within the Store. This approach entails direct liability of actions,
which can act as another deterring factor for malevolent behavior.

Finally, Store failures are never critical: DSAs resort to standard
socket connectivity in case of any unrecoverable error at either the
device or the Store-side.

\subsection{Operational Expenditure Model}

The Store does not impose restrictions to the pricing of modules,
which is left to the discretion of their creators, following other
well-known store paradigms. Nonetheless, the Store offers a standard
way to keep track of the operational resource expenditure per module,
facilitating their pricing.

Certain network resources have a running costs, defined by the Providers.
These can exemplary refer to expended storage, processing time (e.g.,
for involved VMs), served traffic volume or even paid access to external
services, such as Internet monitoring and analysis databases. A module
instance is registered as a user of such resources, for each separate
Provider it employs, using the corresponding Provider-offered interfaces.
The Specialist authoring the module is required to implement a standard
member function, $\texttt{.cost()}$, which should iterate over the
underlying Provider interfaces, summarize and return the running resource
expenditure up to the function callback. Weight functions can then
be freely applied to the reported resource expenditure, obtaining
the total monetary cost.

This approach allows for the dynamic enabling of modules, based on
the user's preferences. For example, a module of robust connectivity
can be enabled via the app only when exchanging sensitive data. Overlay
mirror-paths can then be setup and charged only for the time needed.
Actual payment can be delegated fully to the end-user, without the
Store's mediation in the transaction.

\section{Implementation\label{sec:Implementation}}

A Socket Store prototype has been implemented for the purpose of concept
evaluation. The prototype incorporates the components detailed in
Section~\ref{sec:The-Socket-Store-Arch}, within the AnyLogic platform~\cite{XJTechnologies.2015}.
AnyLogic is a JAVA-based multi-paradigm software development environment,
which facilitates agent-based development. Metrics, Facilities and
Resource/Adapter agents are organized in the platform's palettes,
and the Specialist is offered a GUI enabling: i) the composition of
complex modules, ii) their visualization, iii) their debugging and
run-time management. A central view of instantiated agents and their
environments is readily provided, employing the Agent Management tools
offered by AnyLogic. NSDs are stored in XML format, and can be executed
as such by the Store. DSAs are exported in JAVA $\texttt{.jar}$ portable
format.

The considered network is an SDN environment emulated in MiniNet,
controlled by the POX controller~\cite{mininet}. This is directly
mapped to an agent environment, with Resource Agent types representing:
\begin{itemize}
\item An OpenFlow switch, allowing for inspectable and editable routing
tables.
\item A network link providing read-only access to static (end-points and
capacity) as well as dynamically monitored attributes (packet latency,
transfer rate and load).
\end{itemize}
Resource Agent instances are organized in iterable JAVA collections
by the AnyLogic platform. Moreover, the contained agent management
provides a centralized, graphical representation of all agent instances,
readily yielding the network topology. Thus, the specifics of the
underlying SDN technology are abstracted. The same agent representation
and logic can be ported, e.g., to the Internet, using globally distributed
VMs that form overlay networks.

An Adapter Agent is introduced for this evaluation, coupled with an
example of extended DSA. The K-paths Mirroring Agent (KM) duplicates
a connection and its data over K link-disjoint paths, with approximately
identical attributes. It receives as inputs two host endpoints (IPs
and ports), the integer K and the intended data transfer rate and
latency. A DSA invokes the KM agents passing these arguments. Subsequently,
KM is instantiated and attempts to allocate the K mirror-paths between
the endpoints. If the allocation fails, the DSA is informed and a
regular socket is opened. The DSA may also relay the failure event
to the application business logic to initiate a negotiation instead.
In the latter case, the developer should handle the negotiation logic
as well.
\begin{figure}[t]
\subfloat[{\footnotesize{}\label{fig:emulation}Employed topology and Store
components.}]{\begin{centering}
{\scriptsize{}\includegraphics[width=1\columnwidth]{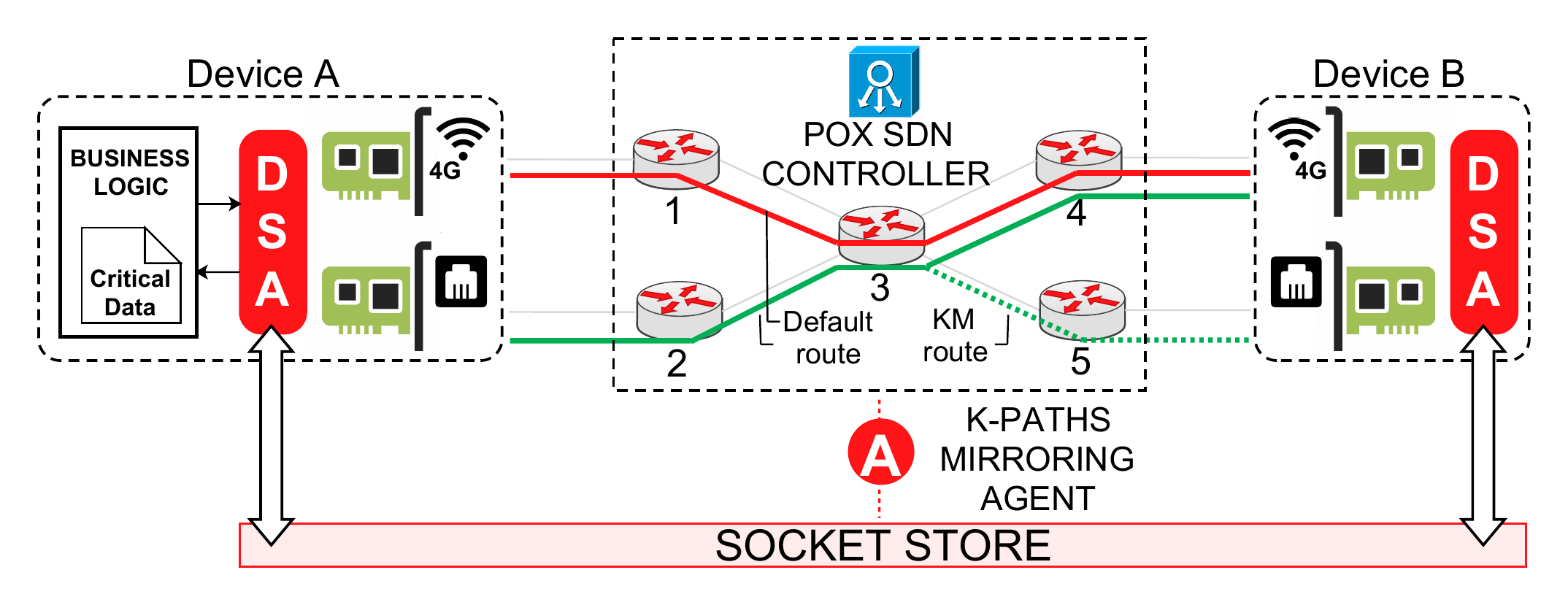}}
\par\end{centering}{\scriptsize \par}
{\scriptsize{}}{\scriptsize \par}}

\subfloat[{\footnotesize{}\label{fig:results}Behavior of the flash-delivery
socket module.}]{\begin{centering}
{\scriptsize{}\includegraphics[clip,width=1\columnwidth]{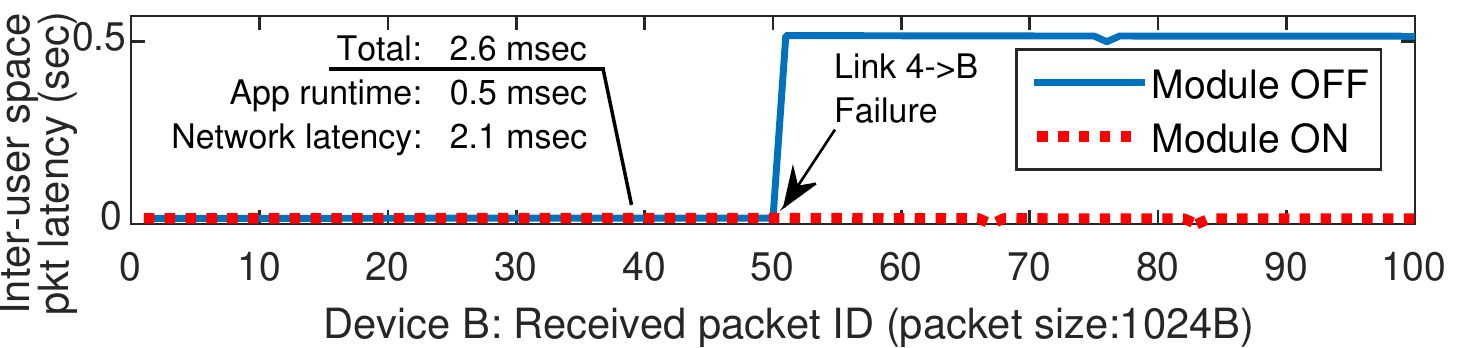}}
\par\end{centering}{\scriptsize \par}
{\scriptsize{}}{\scriptsize \par}}

\caption{{\footnotesize{}\label{fig:Overview-of-the}Overview of the emulation
setup.}}
\end{figure}

We proceed to present the \emph{flash-delivery} module, which employs
the KM Agent. The module usage scenario corresponds to critical data
delivery between two mobile devices over the Internet. Each packet
from a sending \emph{Device A} must reach the recipient \emph{Device
B} with high probability (i.e., using packet replication over link-disjoint
paths) and within a strict deadline (set to $5\,\text{ms}$), without
ACKs or re-transmissions on errors. A single packet loss or deadline
expiry corresponds to failed communication. Flash-trades, critical
VM migrations and RPC signaling are some potential applications~\cite{flashOrder}.

The emulated setup is given in Fig.~\ref{fig:emulation}. Each link
has an effective capacity of $100\,\text{Mbps}$ and $0.5$ ms latency.
We note that the topology comprises $5$ routers due to limitations
of the emulation environment. Specifically, the accuracy of latency
measurements in MiniNet can be low for large topologies~\cite{mininet}.

The network is monitored and controlled by the POX controller located
at node $\#3$, which can deploy/retract paths per flow and maintain
a consistent network-wide state (topology, nodes, link latency).

Each device has two separate physical network cards, and a DSA to
communicate with the Store. At the DSA of Device B, a developer calls
a single line of code, i.e. the exposed $\texttt{bind()}$ function,
which periodically notifies the Store of the current connectivity
details of Device B. At the DSA of the sending side (Device A), a
developer calls a single function, i.e., $\texttt{connect("Device\_B")}$.
The alias $\texttt{"Device\_B"}$ is automatically resolved by the
DSA to the actual connectivity details of B (available at the Store).
Subsequently, the KM agent is instantiated and communicates with the
POX controller on the application's behalf, setting up $2$ mirrored,
link-disjoint paths. (Notice that the default paths may not be link-disjoint,
as shown in Fig.~\ref{fig:emulation}). Finally, the data is sent
on both paths, and device B discards already received packets.

The IPFW tool is used for introducing a sharp latency increase at
link $4\to B$~\cite{IPFW} during the critical data transmission,
as shown in Fig.~\ref{fig:results}. Naturally, when the module is
activated, the data transmission remains successful. \vspace{-5bp}

\section{Discussion and Outlook\label{sec:Summary-and-Outlook}}

The Store offers a novel way for modeling and understanding the interactions
among the various network processes and the applications. Using the
mature, multi-agent approach, the Store promotes the interconnectivity
and re-usability of network modules. Moreover, given that the Store
modules carry a clear objective, the Resource Providers can obtain
a better understanding of the developer's needs, evolving their offered
services accordingly. In this aspect, note that evolving the Internet
protocols and infrastructure presently constitutes a significant research
goal, especially towards improved QoS~\cite{Gupta.2014}. Such paradigms
are top-to-bottom approaches for advancing the Internet: changes to
the Internet core are required, without assuring a monetary profit
for the Network Resource providers first. The Socket Store works from
the complimentary, bottom-to-top direction. It can create a client
base to motivate this Internet core evolution.

Deriving from its compatibility with the current Internet operation,
the Store is not a form of paid prioritization. The Store is a clear
case of outsourced programming: Specialists handle the creation of
modules for a price. This can translate to better Internet service
compared to applications outside the Store. However, this performance
boost will be owed to ``smarter'' network programming, not to unfair
access provision. Additionally, the Store is not a replacement for
existing client-side code libraries, given that basic connectivity
can be sufficient for many applications. Finally, the Store does not
claim an exclusiveness over outsourcing. Freelance Specialists already
offer their services and can freely continue to do so outside any
Store. However, the highly successful cases of mobile app stores have
shown that the added visibility benefits the Specialists. The Socket
Store has the additional prestige potential stemming from its academic
research orientation.

The Store is intended as a unified platform for hosting modular solutions
for a wide variety of application contexts, beyond the described functionality
on the Internet and access networks. For instance, the Store is the
intended interface for an emerging hardware paradigm, the software-defined
metasurfaces (SDMs)~\cite{TheVISORSURFproject.2017}. SDMs are a
novel class of materials with customizable electromagnetic behavior.
They allow for custom steering or absorption of impinging electromagnetic
waves, enabling novel applications in high-resolution imaging and
networking. Modular solutions that describe intended behaviors of
SDMs can be expressed as Store modules, and be activated upon demand
by DSAs. Creating, chaining and multiplexing such functionalities
is a future research goal. \vspace{-5bp}

\section{Conclusion\label{sec:Conclusion}}

The present work introduced the Socket Store, a novel approach for
realizing network-aware applications and application-aware networks.
The Store makes state-of-the-art research knowledge accessible to
mobile/desktop software developers. It motivates appropriately trained
network specialists to publish their work in the form of reusable
software modules. Network providers are encouraged to expose their
infrastructure capabilities as blocks for building advanced modules.
Developers purchase access to a module and enjoy state-of-the-art
networking performance, while minimizing network development time
and its associated expenses. The Socket Store acts as a focal point
for close collaboration among network providers, researchers and developers,
with distinct, fitting roles and benefits.\vspace{-5bp}

\bibliographystyle{IEEEtran}

\end{document}